\documentclass[runningheads]{llncs}
\usepackage{graphicx}
\usepackage{todonotes}
\usepackage{amsmath}
\usepackage{times}
\usepackage{enumitem}
\usepackage{array}
\usepackage{xspace}
\usepackage{multirow}

\newcolumntype{P}[1]{>{\centering\arraybackslash}p{#1}}
\newcolumntype{M}[1]{>{\centering\arraybackslash}m{#1}}

\newcommand{\OR}{\ensuremath{\mathcal{OR}}\xspace}

\begin{document}
%
\title{Query-based Industrial Analytics over Knowledge Graphs with Ontology Reshaping}

\titlerunning{KG-Based Industrial Analytics Enhanced by Ontology Reshaping}
\author{Zhuoxun Zheng\inst{1,2} \and
Baifan Zhou\inst{3} \and
Dongzhuoran Zhou\inst{1,3} \and
Gong Cheng\inst{4} \and
Ernesto Jiménez-Ruiz\inst{5,3}\and
Ahmet Soylu\inst{2} \and
Evgeny Kharlamov\inst{1,3} 
}
\authorrunning{Zheng et al.}
\institute{Bosch Center for Artificial Intelligence, Renningen, Germany \and
Department of Computer Science, Oslo Metropolitan University, Oslo, Norway \and
SIRIUS Centre, University of Oslo, Oslo, Norway \and
State Key Laboratory for Novel Software Technology, Nanjing University, China
\and
Department of Computer Science, City, University of London, London, UK}
\maketitle              
\begin{abstract}
Industrial analytics that includes among others equipment diagnosis and anomaly detection heavily relies on integration of heterogeneous production data. Knowledge Graphs (KGs) as the data format and ontologies as the unified data schemata are a prominent solution that offers high quality data integration and a convenient and standardised way to exchange data and to layer analytical applications over it. However, poor design of ontologies of high degree of mismatch between them and industrial data naturally lead to KGs of low quality that impede the adoption and scalability of industrial analytics. Indeed, such KGs substantially increase the training time of writing queries for users, consume high volume of storage for redundant information, and are hard to maintain and update. 
To address this problem we propose an ontology reshaping approach to transform ontologies into KG schemata that better reflect the underlying data and thus help to construct better KGs. In this poster we present a preliminary discussion of our on-going research,  evaluate our approach with a rich set of SPARQL queries on real-world industry data at Bosch and discuss our findings.

\end{abstract}

\section{Introduction}
\label{sec:intro}
\vspace{-1ex}

Industrial analytics includes among others equipment diagnosis and anomaly detection~\cite{ur2019role}. 
It helps to reduce the downtime of manufacturing equipment, resource consumption, error rates, etc. and aims at enhancing the overall production value-chain which is
one of the key goals of 
Industry 4.0~\cite{kagermann2015change,zhou2018comparison}.
Industrial analytics heavily relies on integration of heterogeneous production data. 
Knowledge Graphs (KGs) as the data format for integration and ontologies as the unified data schemata are a prominent solution that offers not only a high quality data integration~\cite{DBLP:journals/internet/HorrocksGKW16} but also a convenient and standardised way to exchange data and to layer analytical applications over it~\cite{zhou2022ontoreshape,zou2020survey}.\looseness=-1

However, poor design of ontologies or high degree of mismatch between them and industrial data,
e.g., when the ontology is designed to reflect  the general domain of knowledge or exported, rather than data particularities~\cite{yahya2022towards}
naturally lead to KGs of low quality that impede the adoption and scalability of industrial analytics. 
Indeed, such KGs often have deep structure or have many blank nodes thus
they are sparse and consume a high volume of storage for redundant information, and are hard to maintain and update.
Moreover, accessing such data requires long and cumbersome SPARQL queries that are counterintuitive and this 
substantially increases the training time of users such as engineers~\cite{SoyluKZJGSHSBLH18}
in writing such queries.\looseness=-1

In order to address this problem, we propose an ontology reshaping approach to transform ontologies into KG schemata that better reflect the given underlying industrial data and thus help to construct better KGs. In this poster, we present a preliminary discussion of our on-going research, evaluate our approach with a rich set of SPARQL queries on real-world industry data at Bosch and discuss the our findings.


\section{Our Approach}
\label{sec:approach}
\vspace{-1ex}

\begin{figure}[t]
\vspace{-3.5ex}
\centering
\includegraphics[width=\textwidth]{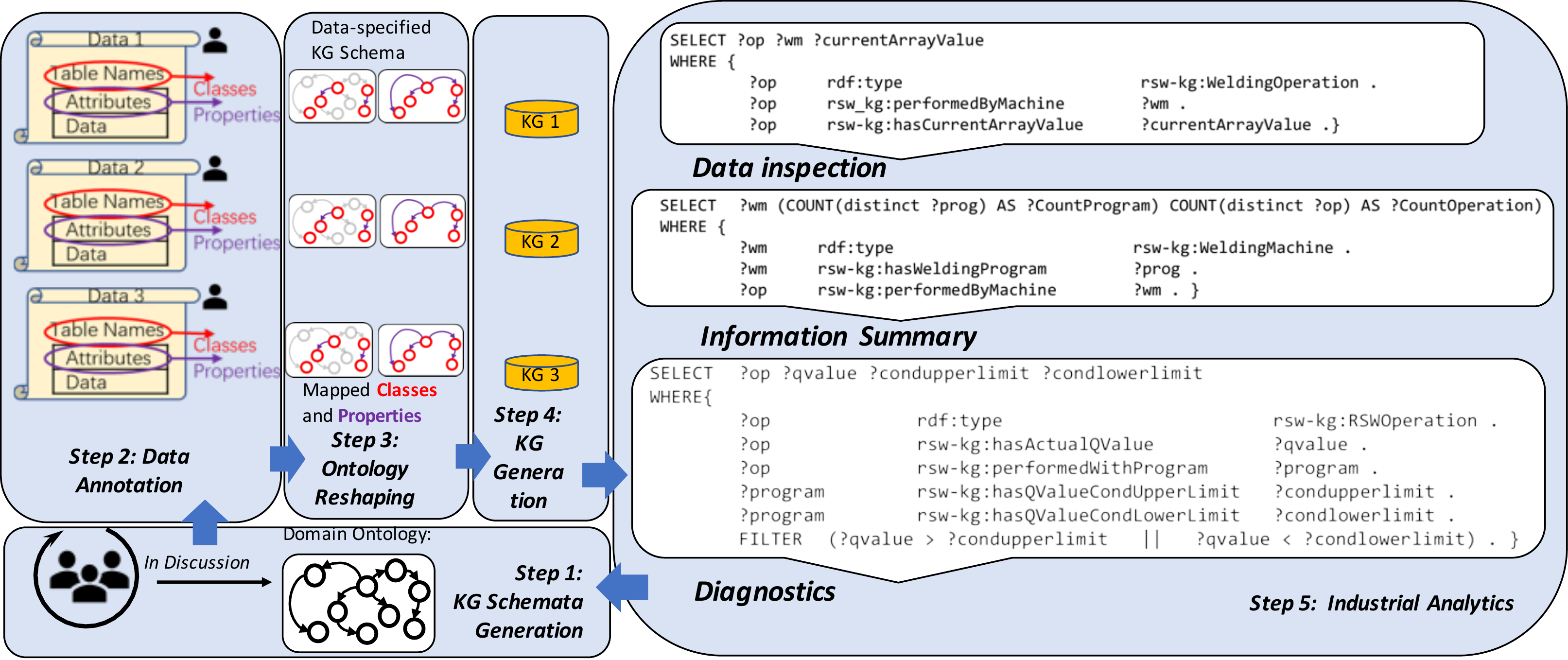}
 \vspace{-6ex}
\caption{Our workflow of enhancing industrial analytics with ontology reshaping (\OR)}
\label{fig:Approach}
\vspace{-4ex}
\end{figure}

 We now describe our approach of enhancing KG generation with \OR that consists of five steps and summarised in 
 Fig.~\ref{fig:Approach}.

\smallskip
\noindent \textbf{KG Schemata Generation} (Step 1, left-bottom part of Fig.~\ref{fig:Approach}). In this step, we adopt the approach of generating KG schemata by extending domain ontologies from upper level ontologies~\cite{zhou2021semml,DBLP:conf/semweb/SvetashovaZPSSM20}.
The users, i.e. domain experts, participate heavily in ontology extension. They have intensive discussions  and create good quality ontologies that reflect the general domain knowledge.

\smallskip
\noindent \textbf{Data Annotation} (Step 2, Fig.~\ref{fig:Approach})
In this step experts annotate
 heterogeneous
 industrial 
raw data  with diverse formats 
 and structures
 collected from production  
with unified terms in the domain ontology created in Step 1. Here we consider the raw data  in the format of relational tables. The table names are mapped to classes 
 in the ontology 
and attribute names to properties. In some complex cases, attribute names with endings like ``ID'' or ``NAME''  are elevated to classes according to users' inputs.\looseness=-1

\smallskip
\noindent \textbf{Ontology Reshaping} (Step 3, Fig.~\ref{fig:Approach})
In this step our ontology reshaping algorithm \OR takes a knowle\-dge-oriented domain ontology, a raw dataset, a mapping between them, and some optional user heuristics as the input and outputs a data-orientated ontology, which serves as the KG schemata. 
The resulting ontology is a (compact) version of the original one and essentially consists of 
(1) all corresponding information (table names and attributes) from the raw data and 
(2) other essential connecting elements, which are partially from the original ontology and partially from users, to attain some optimality defined by user heuristics, efficiency, simplicity, etc.
We adopt our \OR algorithm~\cite{ijckg2021ontoreshaping}.
In the nutshell \OR firstly
selects a subset of nodes and edges in the domain ontology~\cite{DBLP:conf/semweb/SvetashovaZSPK20},
creating its sub-graph, which is a sparse sub-graph that consists
of many disconnected small graph fragments; secondly, it
extends the sub-graph to a KG schema with the help of two
sources of information: 
(a) retaining other nodes and edges of
the domain ontology to preserve part of its knowledge, 
(b) some
optional information given by users (welding experts) that help to connect the fragments in the sub-graph.


\smallskip
\noindent \textbf{KG Generation} (Step 4, Fig.~\ref{fig:Approach})
In this step the KG schemata are populated in the ETL fashion with the actual data based on the annotated table names and attributes in the relational table to generate the KGs. The generated KGs are data-oriented and (often) more compact.

\smallskip
\noindent \textbf{Industrial Analytics} (Step 5, Fig.~\ref{fig:Approach})
In this step, we layer industrial analytics upon the generated KG.
In this paper, we consider three types of SPARQL queries written by engineers or generated by users' keywords inputs. The queries account for retrieving or summarising specific information of interest from huge datasets that come from running production lines, and to perform basic diagnostics over them.
The details follow in the next section. 
During the application, the user feedback is constantly collected and the workflow can go back to Step 1 
and the whole process restarts.
For example, after the application stage, users may realise they can scale the system to more tasks and thus go back to extend the domain ontology for more tasks.\looseness=-1

\smallskip
\noindent\textbf{Related Work.}
The most related work to ours are the one on ontology modularisation \cite{doran2006ontology}, summarisation~\cite{DBLP:conf/www/ZhangCQ07,DBLP:journals/ijsc/PouriyehALCAAMK19}, and summarisation forgetting.
They focus on selecting subsets of ontologies, but they do not address the data particularity issue.
We thus propose to rely on ontology reshaping (\OR), which transforms domain ontologies to its often compacter versions that reflect more  data specificities. The resulting KGs will contain no blank-nodes and become less deep, and thus the queries will become less deep and more user-friendly. 



\section{Evaluation with Industrial Dataset}
\label{sec:application}
\vspace{-1ex}

We now present evaluation of our approach  with a real industrial dataset.














\smallskip
\noindent \textbf{Data Description.}
The dataset $D$ is collected from a German factory, in which reside production lines that consist of 27 welding machines of an impactful automated welding process widely applied in automotive industry: the resistance spot welding~\cite{DBLP:journals/jim/ZhouPRKM22,DBLP:conf/jist/ZhouZCS0K21}. $D$ contains a high number of welding operation records and a series of welding sensor measurements. These data account for 1000 welding operations, estimated to be related to 100 cars.  In this work,
we select a section of $D$ for discussion, which contains 4.315 million records and 176 attributes. 
The knowledge-oriented domain ontology $O$ is generated by welding experts and contains
206 classes, 203 object properties, and 191 datatype properties.
The mapping maps all 176 attributes in $D$ to classes in $O$.

\smallskip
\noindent \textbf{Query Description.}
We consider SPARQL queries
of three types as follows:
\textit{Type~I: Data inspection}, where the experts need to inspect these data for generating a first handful of insights.
    The desired answer to such query is a listing of some attributes. An example see Step 5 in Fig.~\ref{fig:Approach}, which will return all arrays of currents with the corresponding operation-names and machine-IDs.
\textit{Type~II: Information summary}, where the welding experts need to gain overview information of arbitrarily selected datasets,
   e.g. how many different programs does every welding machine perform (Step 5 in Fig.~\ref{fig:Approach})?  
%
\textit{Type~III: Diagnostics}, where the welding experts need to perform various
diagnostic tasks, such as detecting abnormal machines, operations, etc. Moreover, the users would also like to find the surroundings of the abnormalities, to figure out what happened near the abnormalities, so that they can better understand 
for root-causes. 
    One example for this kind of queries would be: Where are the abnormal welding operations whose quality indicators exceed the conditional tolerance limit? (Step 5 in Fig.~\ref{fig:Approach})


\smallskip
\noindent \textbf{Experiment Design.}
To test our approach, we randomly sub-sample $D$ to 6 sub-datasets (Set 1-6 in Table \ref{tab:evaluation}). Each set contains a subset of the attributes of $D$, reflecting different data complexity. The numbers of attributes in the subsets increase by twenty each time, from 20 to 120. We repeat the sub-sampling for each subset 10 times to decrease the randomness.
We compare our approach with a \textit{baseline} of KG generation
without ontology reshaping, which is a naive approach to
use the 
domain ontology directly as the KG schema.
This work considers
729 queries and 324, 189, 216 queries for query Type I, II, III, respectively. 
The evaluation metrics are set as the average and maximal query depth, which characterises the number of edges to connect two nodes in the query graph via the shortest path (the query is essentially also a graph with variable nodes).

\smallskip
\noindent \textbf{Results and Discussion.}
The results (Table~\ref{tab:evaluation}) show that for retrieving the same answers, the queries are significantly simplified with our approach:
the query depths are reduced by about 2 for both its average and maximum. This indicates the generated KG 
becomes more practical,
because shorter queries are needed 
to get the same information. In addition, we also observe that the KGs after Onto-Reshape become more efficient and simpler: their generation
becomes 7 to 8 times faster, the number of entities are reduced to  1/2 to 1/6 of the
baseline, storage space  to 2/3, and the number of blank nodes
to zero.

\begin{table*}[t]
 \vspace{-2ex}
\renewcommand*{\arraystretch}{1.1}
	\setlength{\tabcolsep}{0.55mm}
	\center
	\caption{{Our approach enhanced by ontology reshaping (Onto-Reshape) outperforms the baseline significantly in terms of query simplicity, avg.: average, max: maximum\looseness=-1}}
	\vspace{-2ex}
        \begin{tabular}{M{1.2cm}|P{3cm}|P{1cm}|P{1cm}|P{1cm}|P{1cm}|P{1cm}|P{1cm}}
        \hline
\multicolumn{2}{c|}{Subset}     &  Set 1  &  Set 2  &  Set 3  &  Set 4  &  Set 5  &  Set 6  \\\hline 
raw data  &  \#attributes  & 20 & 40 & 60 & 80 & 100 & 120  \\ \hline
\multirow{2}{=}{Baseline} 
  &  avg. query depth  & 4.2 & 4.4 & 4.3 & 4.3 & 4.2 & 4.3  \\ \cline{2-8}
  &  max. query depth  & 5.0 & 5.0 & 5.0 & 5.0 & 5.0 & 5.0 \\ \cline{2-8} \hline
\multirow{2}{=}{Onto Reshape}    &  avg. query depth   & 2.3 & 2.5 & 2.6 & 2.7 & 3.0 & 3.1 \\ \cline{2-8}
  &  max. query depth   & 3.0 & 4.0 & 3.0 & 4.0 & 3.0 & 4.0   \\ \hline
        \end{tabular}
    \label{tab:evaluation}
    \vspace{-2ex}
\end{table*}

\section{Conclusion and Outlook}
\label{sec:conclusion}
\vspace{-1ex}


In this paper, we present our ongoing research of knowledge graph-based industrial query analytics at Bosch. 
We have preliminary results that show the benefit of our approach for industrial analytics. 
Our work falls into the big picture of KG-based industrial applications~\cite{zhou2022exploiting}. 
As future work we will investigate transforming queries across different KG schemata, and study more query properties.\looseness=-1



\medskip \noindent \textbf{{Acknowledgements.}}
The work was partially supported by the H2020 projects Dome 4.0 (Grant Agreement No. 953163), OntoCommons (Grant Agreement No. 958371),
and DataCloud (Grant Agreement No. 101016835) and the SIRIUS Centre,
Norwegian Research Council project number 237898.

\bibliographystyle{elsarticle-num}
\bibliography{eswc-orautokg}

\everypar{\looseness=-1}

\end{document}